\documentclass[12pt]{iopart}

\usepackage{dcolumn}
\usepackage{bm}
\usepackage{graphicx}
\usepackage{color}
\usepackage{subfigure}
\usepackage{hyperref}
\usepackage{latexsym}
\usepackage{amsthm}
\usepackage{amssymb}
\usepackage{braket}

\usepackage{cite}
\DeclareGraphicsExtensions{.jpg,.pdf, .mps, .png, .eps, .ps, .EPS,.gif}

\newcommand{\Avg}[1]{\left\langle{#1}\right\rangle}

\begin{document}
\def\be{\begin{equation}}
\def\ee{\end{equation}}

\def\bc{\begin{center}}
\def\ec{\end{center}}
\def\bea{\begin{eqnarray}}
\def\eea{\end{eqnarray}}

\def\ie{\textit{i.e.}}
\def\etal{\textit{et al.}}
\def\m{\vec{m}}
\def\G{\mathcal{G}}

\newcommand{\gin }[1]{{\bf\color{cyan}#1}}

\title[The mass  of simple and higher-order networks ]{The  mass  of simple and higher-order networks }

\author{Ginestra Bianconi$^{1,2}$}

\address{$^1$ School of Mathematical Sciences, Queen Mary University of London, E1 4NS London, United Kingdom\\
$^2$ Alan Turing Institute, The British Library, London, United Kingdom}

\ead{ginestra.bianconi@gmail.com}
\vspace{10pt}
\begin{indented}
\item[]
\end{indented}

\begin{abstract}
We propose a theoretical framework that explains how the mass of simple and higher-order networks emerges from their topology and geometry. We use the discrete topological Dirac operator to define an action for a massless self-interacting topological Dirac field inspired by the Nambu-Jona Lasinio model.  {The mass of the network is strictly speaking the mass of this topological Dirac field defined on the network; it  results from the chiral symmetry breaking of the model and satisfies a self-consistent gap equation.} Interestingly, it is shown that the mass of a network depends on its spectral properties, topology, and geometry.
Due to the breaking of the matter-antimatter symmetry observed for the harmonic modes of the discrete topological Dirac operator, two possible definitions of the network mass can be given. For both possible definitions, the mass of the network comes from a gap equation with the difference among the two definitions encoded in the value of the bare mass. Indeed, the bare mass can be determined either by the Betti number $\beta_0$ or by the Betti number $\beta_1$ of the network.
We provide numerical results on the mass of different networks, including random graphs, scale-free, and real weighted collaboration networks. We also discuss the generalization of these results to higher-order networks, defining the mass of simplicial complexes.  {The observed dependence of the mass of the considered topological Dirac field with the topology and geometry of the network could lead to interesting physics in the scenario in which  the considered Dirac field is coupled with a dynamical evolution of the underlying network structure.}
\end{abstract}

%
%
%
%
%
\section{Introduction}
Recently, growing attention is addressed to the discrete topological Dirac operator~\cite{bianconi2021topological, bianconi2023dirac}, originally introduced in non-commutative geometry~\cite{davies1993analysis, krajewski1998classification, paschke1998discrete, requardt2002dirac, majid2023dirac}, and then used in quantum graphs~\cite{post2009first, hinz2013dirac, anne2015gauss, athmouni2021magnetic, parra2017spectral, bolte2003spectral, bolte2003spin, kramar2020linear} and in network theory~\cite{baccini2022weighted, giambagli2022diffusion, calmon2022dirac, calmon2023dirac, calmon2023local, wee2023persistent}.
On a network, the discrete topological Dirac operator is defined over topological spinors, which are the direct sum of $0$-cochain and $1$-cochains. The Dirac operator in its basic form is defined as $d+d^{\star}$, where $d$ here indicates the exterior derivative of the network. This operator is key to defining the discrete topological Dirac equation~\cite{bianconi2021topological}.
This equation has a big difference with the continuous one, as the eigenstates corresponding to energies $E=m$ of particles and antiparticles do not satisfy the matter-antimatter symmetry~\cite{bianconi2021topological}. In particular, the eigenstates at energy $E=m$ of matter and antimatter have different degeneracies given by the Betti numbers $\beta_0$ and $\beta_1$ of the network. It follows that as soon as the network is not a closed chain, their degeneracy is different. The Dirac operator can also be coupled to an algebra~\cite{bianconi2021topological, bianconi2023dirac} in order to provide a discrete topological Dirac equation that corresponds to the continuous $(d+1)$-dimensional Dirac equation.

In this work, we define the mass of simple and higher-order networks drawing inspiration from the Nambu-Jona-Lasinio model~\cite{nambu1961dynamical}. We show that a massless discrete Dirac field obeying the topological Dirac equation on a network can acquire a mass as a result of chiral symmetry breaking. Interestingly, we see that this mass depends on both the topology and the geometry of the network, as it can be defined for both weighted and unweighted networks. As in the original Nambu-Jona-Lasinio model~\cite{nambu1961dynamical}, here we demonstrate that the mass of a network satisfies a self-consistent gap equation, which is analogous to the gap equation in BCS theory of superconductivity~\cite{bardeen1957microscopic}. Moreover, we demonstrate that the defined mass depends on the network's spectral properties, its metric, and its topology (through its Betti numbers). As mentioned before, the gap equation of a network shares significant similarities with the gap equation obtained in the Nambu-Jona-Lasinio model~\cite{nambu1961dynamical}, used as an effective theory in quantum chromodynamics~\cite{hatsuda1994qcd, klevansky1992nambu, buballa2015inhomogeneous}. However, there are also important differences. The first one is that the small bare quark mass used in quantum chromodynamics is in our framework dictated by the network's Betti numbers, hence giving an important topological character to the bare mass value. The other important difference is that there is no dependence on the cutoff value, as the network is discrete.

Furthermore, another important difference exists with the continuous Nambu-Jona-Lasinio model. Since the topological Dirac equation defines the topological spinor on nodes and links, and since in general - with the exception of the one-dimensional closed chain - there is no duality between nodes and links, then one has, in principle, the liberty to choose in which of the two sectors in the classical limit the particles and the antiparticles live. This property of the topological Dirac equation implies that there are two possible definitions of the Dirac mass of the network, one depending on the Betti number $\beta_0$ and one depending on the Betti number $\beta_1$.

Here we show that the resulting mass of random graphs~\cite{ER} and the scale-free Barabasi-Albert (BA) model~\cite{BA} depends on their average degree, which notably determines their Betti numbers and their spectral properties. The investigation of the mass of the Bianconi-Barabasi (BB) trees displaying the so-called Bose-Einstein condensation in complex networks~\cite{bianconi2001bose} provides evidence that even if the Betti numbers do not change, the mass of the network can strongly depend on the network structural properties. Finally, by comparing the mass of weighted and unweighted collaboration networks, we provide evidence that the mass of a network can also depend on the network geometry through the metric degrees of freedom of the network. The paper concludes with the generalization of this approach leading to the definition of the mass of higher-order networks~\cite{bianconi2021higher, battiston2021physics, battiston2022higher}. In this context, we demonstrate that for simplicial complexes of any dimension, there are always two possible definitions of the mass, which coincide only when the Euler characteristic of the simplicial complex is zero.

The observed dependence of the mass of the network with its topology and geometry could opens new scenarios in the case in which we assume that the network structure is not fixed in time but evolves together with the topological Dirac field. Considering the coupled dynamics of the topological Dirac field with the metric and topological degree of freedom of the network could be the subject of further research.

The proposed theory provides new perspectives for research at the interface between network theory, quantum gravity~\cite{Rovelli, CDT, calcagni2013probing, calcagni2013laplacians, Lionni, Dario, Astrid, Codello, Oriti, baez1996spin,benvenuti2007counting}, supersymmetry~\cite{witten1982supersymmetry, frezzotti2001lattice}, and lattice gauge theories~\cite{rothe2012lattice, dalmonte2016lattice, banuls2020simulating, aidelsburger2018artificial, surace2020lattice}. Despite here drawing direct connections across these fields is beyond the scope of this work, the relations among these different fields could be the subject of future research works.

This work is structured as follows. In Sec. \ref{sec:simple}, we define the theory leading to the definition of the mass of a generic network, and we provide evidence that the network mass depends on the topology and the geometry of the network. In Sec. \ref{sec:higher}, we extend this framework to define the mass of higher-order simplicial complexes. Finally, we provide the concluding remarks.

\section{The mass of simple networks}
\label{sec:simple}
In this Section we provide the theoretical framework that gives rise to the definition of the {\em mass of a network}. As we will see this term strictly speaking indicates the mass of a topological self-interacting Dirac field defined on a network which  emerges from chiral symmetry breaking.

\subsection{Simple network, exterior derivative and its dual}
As background material for this paper here we introduce some basic notions of algebraic topology including boundary and coboundary matrices and Hodge Laplacians~\cite{eckmann1944harmonische,horak2013spectra,bianconi2021higher}.

Let us consider a simple network (also called graph)  $G=(V,\tilde{E})$ where $V$ is the set of  $N_0$ nodes, and $\tilde{E}$ is the set of  $N_1$ links.
Through this work we will assume that links are oriented, with an orientation induced by the node labels so that the link $[i,j]$ is oriented positively if $i<j$. 
A $0$-cochain $f\in C^0$ can be encoded by the vector ${\bf f}$ taking  complex values on each node $i$ of the network, i.e. $f_{[i]}\in \mathbb{C}$.
Similarly a $1$-cochain $g\in C^1$ can be encoded into a vector ${\bf g}$ taking complex values on each (positively oriented) link $\ell={[i,j]}$ of the network, i.e. $g_{\ell}\in \mathbb{C}$.
The exterior derivative $\delta_1:C^{0}\to C^{1}$ defines the discrete gradient and acts on the $0$-cochains so that if $g=\delta_1f$ then 
\bea
g_{[i,j]}=f_{[j]}-f_{[i]}.
\label{d}
\eea
This equation can be written in matrix form as  
\bea
{\bf g}=\bar{\bf B}_{[1]}^U{\bf f},
\eea
where $\bar{\bf B}_{[1]}^U$ indicates the $N_1\times N_0$ unweighted coboundary matrix.
This definition of the discrete gradient can be further enriched by considering a weighted network where each link $[i,j]$ is associated to a real and positive weight $w_{[i,j]}$ and each node is associated to a real and positive weight $s_{[i]}$.
In this setting Eq.(\ref{d}) becomes 
\bea
g_{[i,j]}=\sqrt{w_{[i,j]}}\left(\frac{f_{[j]}}{\sqrt{s_{[j]}}}-\frac{f_{[i]}}{\sqrt{s_{[i]}}}\right).
\label{dw}
\eea
Let us define the  diagonal $N_0\times N_0$ metric matrix ${\bf G}_{[0]}$ and the diagonal $N_1\times N_1$ metric matrix ${\bf G}_{[1]}$ of diagonal elements  
\bea
{\bf G}_{[0]}([i][i])=s_{[i]}^{-1}\nonumber \\
{\bf G}_{[1]}([i,j][i,j])=w_{[i,j]}^{-1}.
\eea
 In matrix form Eq. (\ref{dw}) reads 
\bea
{\bf g}=\bar{\bf B}_{[1]}{\bf f},
\eea
where the $N_1\times N_0$ coboundary matrix $\bar{\bf B}_{[1]}$ is given by 
\bea
\bar{\bf B}_{[1]}={\bf G}_{[1]}^{-1/2}\bar{\bf B}_{[1]}^U{\bf G}_{[0]}^{1/2}.
\eea
We can then introduce a scalar product between cochains here taken to be the $L^2$ norm.  Therefore if $f$ and $\tilde{f}$ indicate two different $0$-cochains and if $g$ and $\tilde{g}$ indicate two different $1$-cochains, then their scalar product is defined as 
\bea
\Avg{f,\tilde{f}}=\sum_i f_{[i]}\tilde{f}_{[i]},\quad
\Avg{g,\tilde{g}}=\sum_{\ell=[i,j]} g_{[i,j]}\tilde{g}_{[i,j]}.
\eea 
The definition of the scalar product between cochains allows us to define the adjont operator  $\delta_1^{\star}:C^1\to C^0$ of the discrete gradient as the operator satisfying 
\bea
\Avg{g,\delta_1f}=\Avg{\delta_1^{\star}g,f},
\eea
for any $f\in C^{0}$ and $g\in C^1$.
It is straightforward to see that this definition implies that this operator can be identified as the discrete divergence and that  if $g=\delta_1^{\star}f$ then ${\bf g}=\bar{\bf B}^{\top}{\bf f}$. We will call $\bar{\bf B}^{\top}$ the boundary matrix of the network.
By multiplying the boundary matrix with the coboundary matrix one obtains the Laplacians operators ${\bf L}_{[0]}=\bar{\bf B}^{\top}\bar{\bf B}$ and ${\bf L}_{[1]}^{down}=\bar{\bf B}\bar{\bf B}^{\top}$ that describe respectively diffusion from nodes to nodes through links and from links to links through nodes. 
The $0$-Hodge Laplacian ${\bf L}_{[0]}$ is the symmetrized weighted graph Laplacian extensively used in spectral graph theory \cite{chung1997spectral} reducing to the graph Laplacian when the metric matrices are identities.
The $1$-Hodge Laplacian is given by ${\bf L}_{[1]}={\bf L}_{[1]}^{down}$.
Both Hodge Laplacians are semi-definite positive and the dimension of their kernel is given by the corresponding Betti number, i.e.
\bea
\mbox{dim} \ \mbox{ker} {\bf L}_{[n]}=\beta_n
\eea
where on a connected network $\beta_0=1$ and $\beta_1=N_1-N_0$,  {i.e. the Betti  number $\beta_0$ is equal to the number of connected components, while the Betti number $\beta_1$ is equal to the number of independent cycles of the network.}

\subsection{Topological Dirac operator}

We will consider a topological spinor $\bm\psi\in C^0\oplus C^1$ defined on nodes, links and  given by the direct sum of $0,1$  cochains, which in matrix form reads
\bea
\bm\psi=\left(\begin{array}{c} \bm\phi\\ \bm\xi \end{array}\right),
\eea
where $\bm\phi\in C^0, \bm \xi\in C^1$ or  equivalently
\bea
\bm \phi=\left(\begin{array}{c} \phi_1\\ \phi_2\\\vdots \\\phi_{N_0}\end{array}\right), & \bm \xi=\left(\begin{array}{c} \xi_{\ell_1}\\ \xi_{\ell_2}\\\vdots\\ \xi_{\ell_{N_1}}\end{array}\right).
\eea
On top of the topological spinor we define the discrete topological Dirac operator ${\bf D}:C^0\oplus C^1 \to C^0 \oplus C^1 $ that maps topological spinors to topological spinors.
The Dirac operator is defined in matrix form as
\bea
\mathbf{D}&=&\left(\begin{array}{cc}0& {\bf \bar{B}}_{[1]}^{\top}\\{\bf \bar{B}}_{[1]}& 0\end{array}\right).
\label{dirac_simple}
\eea
The Dirac operator is the ``square root" of the  Laplacian as we have 
\bea
{\bf D}^2={\bf\mathcal{L}}=\left(\begin{array}{cc} {\bf {L}}_{[0]}& 0\\0&{\bf {L}}_{[1]}\end{array}\right).
\eea
Since the $0$-Hodge Laplacian and the $1$-Hodge Laplacian of a network are  {isospectral}, the eigenvalues $\lambda$ of the Dirac operator are given by 
\bea
\lambda=\pm \sqrt{\mu}
\eea
where $\mu$ indicates any eigenvalue in the spectrum of one of the graph Laplacians.
Notably, this implies that the Dirac operator has no definite sign.
The eigenvectors  corresponding to the eigenvalues with the same absolute value, are related by chirality.
In order to see this let us define the gamma matrix $\bm\gamma_0$ as
\bea
\bm \gamma_0=\epsilon\left(\begin{array}{cc}{\bf I}& 0\\0& -{\bf I}\end{array}\right),
\label{gamma_0}
\eea
with $\epsilon\in \{1,-1\}$.
From the definition of the Dirac operator given in Eq. (\ref{dirac_simple}), it is straightforward to see that the anticommutator between the Dirac ${\bf D}$ operator and the gamma matrix $\bm\gamma_0$ vanishes, i.e.
\bea
\{\bm\gamma_0,{\bf D}\}=0.
\eea
This implies that if $\bm\psi=(\bm\phi^{\top},\bm\xi^{\top})^{\top}$ is an eigenvector of the Dirac operator with eigenvalue $\lambda>0$, then $\bm\psi=(\bm\phi^{\top},-\bm\xi^{\top})^{\top}$ is an eigenvector of the Dirac operator with eigenvalue $\lambda^{\prime}=-\lambda$.
Note however that this chiral symmetry only involves the non-harmonic eigenvectors of the topological Dirac operator.
On the contrary the harmonic eigenvectors of the topological Dirac equation can be chosen to be either localized on nodes $\bm\psi=(\bm\phi_0^{\top},{\bf 0}^{\top})^{\top}$ or localized on links  $\bm\psi=({\bf 0}^{\top},\bm\xi_0^{\top})^{\top}$. The multiplicity of the harmonic eigenvectors of the Dirac operator that are localized on the nodes is given by the Betti number $\beta_0$, while the multiplicity of the harmonic eigenvectors of the Dirac operator that are localized on the links is given by the Betti number $\beta_1$. Thus the dimension of this two spaces is different as long as the network Euler number $\chi=\beta_1-\beta_0$ is non-zero, i.e. as long as the network departs from a closed chain.

\subsection{Topological Dirac field theory}

The Dirac equation
\cite{dirac} describes spin $1/2$ particles and has had profound ramification in quantum mechanics, topology and condensed matter \cite{thaller2013dirac,pais2005paul,cuevas2015solitary,cuevas2018solitary}.
Having introduced the discrete topological Dirac operator we can now discuss the topological Dirac equation \cite{bianconi2021topological} which provides a discrete counterpart to the continuous Dirac equation~\cite{dirac} emphasizing their relations and their differences. Moreover here we  implement the second quantization of the topological Dirac equation hence  proposing its corresponding topological Dirac field theory.
Our notation and our conventions are the natural extension to the discrete and topological setting of the notations  adopted in Ref.
 \cite{lebellac} for the continuous Dirac field theory.
 
 We consider the action $\mathcal{S}$ given by 
\bea
\mathcal{S}=\bar{\bm\psi}[\bm\gamma_0({\rm{i}}\partial_t-{\bf D})-m_0]\bm\psi
\label{S_simple}
\eea
where $\bar{\bm\psi}=\bm\psi^{\dag}\bm\gamma_0$, and $m_0>0$ is the bare mass.  The corresponding path integral is defined as
\bea
Z=\int \mathcal{D}\bm\psi\int \mathcal{D}\bar{\bm\psi}e^{\rm{i}\mathcal{S}}.
\eea
The equation of motion provides the topological Dirac equation \cite{bianconi2021topological},
\bea
{\rm{i}}\bm{\gamma}_0\partial_t{\bm\psi}=\bm{\gamma}_0{\bf D}{\bm\psi}+ m_0 {\bm\psi}.
\eea
By using the definition of the gamma matrix $\bm\gamma_0$ given by Eq. (\ref{gamma_0}) and defining $m$ as $m=\epsilon m_0$, we can show that the eigenstates of the topological Dirac equation with energy $E$ can be written as 
\bea
E\bm\phi=  {\bf \bar{B}}_{[1]}^{\top} \bm\xi+ m\bm\phi,\nonumber \\
E\bm\xi=  {\bf \bar{B}}_{[1]} \bm \phi-m\bm\xi.
\eea
By collecting the terms proportional to $\bm\phi$ and to $\bm\xi$ we obtain
\bea
(E-m)\bm\phi&=& {\bf \bar{B}}_{[1]}^{\top} \bm\xi,\nonumber \\
(E+m)\bm\xi&=&{\bf \bar{B}}_{[1]} \bm\phi.
\eea
Finally with simple manipulations we get
\bea
(E+m)(E-m)\bm\phi={\bf \bar{B}}_{[1]}^{\top}{\bf \bar{B}}_{[1]} \bm\phi={\bf L}_{[0]}\bm\phi,\nonumber\\
(E+m)(E-m)\bm\xi={\bf \bar{B}}_{[1]}{\bf \bar{B}}_{[1]}^{\top} \bm\xi={\bf L}_{[1]}\bm\xi,
\eea
which implies that $\bm\phi$ is proportional to the eigenvalue of the graph Laplacian ${\bf L}_{[0]}$ with eigenvalue $\mu=\lambda^2$ and $\bm\xi$ is proportional to the eigenvector of ${\bf L}_{[1]}$ with eigenvalue $\mu=\lambda^2$ where $E$ and $\lambda$ are related by the relativistic dispersion relation 
\bea
E^2=\lambda^2+m^2.
\eea
Let us  define ${\bf w}_{\lambda}$ and $\hat{\bf w}_{\lambda}$ as the left and right singular vector of $\bar{\bf B}_{[1]}$ respectively both corresponding to its singular value $\lambda$. Moreover let us assume for the time being that $\epsilon=1$ and hence $m>0$.
The eigenstates of the Dirac equation can be distinguished according the the value of the sign of the energy $E$.
In particular we will indicate with ${\bf u}_{\lambda}$ the eigenstates corresponding to positive energy $E=E_\lambda=\sqrt{\lambda^2+m^2}$ while we will indicate with ${\bf v}_{\lambda}$ the eigenstates corresponding to energy $E=-E_{\lambda}$.
These eigenstates  can be written as 
\bea
{\bf u}_{\lambda}=\frac{1}{\sqrt{E_\lambda+m}}\left(\begin{array}{c} (E_{\lambda}+m){\bf w}_{\lambda}\\ {\lambda}\hat{\bf w}_{\lambda}\end{array}\right),\nonumber \\
{\bf v}_{\lambda}=\frac{1}{\sqrt{E_{\lambda}+m}}\left(\begin{array}{c} -\lambda{\bf w}_{\lambda}\\ (E_{\lambda}+m)\hat{\bf w}_{\lambda}\end{array}\right),
\eea
where  we have adopted the normalization
\bea
\bar{{\bf u}}_{\lambda}{\bf u}_{\lambda}=2m,\quad
\bar{{\bf v}}_{\lambda}{\bf v}_{\lambda}=-2m,\nonumber \\
\bar{{\bf u}}_{\lambda}\bm\gamma_0{\bf v}_{\lambda}=0,\quad
\bar{{\bf v}}_{\lambda}\bm\gamma_0{\bf u}_{\lambda}=0.
\label{uno}
\eea
Note that with this normalization we also have 
\bea
\bar{{\bf u}}_{\lambda}\bm\gamma_0{\bf u}_{\lambda}=2E_{\lambda},\quad
\bar{{\bf v}}_{\lambda}\bm\gamma_0{\bf v}_{\lambda}=2E_{\lambda},\nonumber \\
\bar{{\bf u}}_{\lambda}{\bf v}_{-\lambda}=0,\quad
\bar{{\bf v}}_{-\lambda}{\bf u}_{\lambda}=0.
\label{due}
\eea

 {Now we are in the position to see how changing $\epsilon\to -\epsilon$ and therefore $m\to -m$ effectively change the  eigenstates ${\bf u}_{\lambda}$ corresponding to positive energy states and the  eigenstates ${\bf v}_{\lambda}$ corresponding to negative energy. This transformation will be related to charge conjugation.
Indeed for $\lambda\neq 0$ the eigenstates change as  
\bea
{\bf u}_{\lambda}\to \tilde{\bf v}_{\lambda}={\bf v}_{-\lambda}=-{\bm\gamma}_0 {\bf v}_{\lambda},\quad
{\bf v}_{\lambda}\to \tilde{\bf u}_{\lambda}=-{\bf u}_{-\lambda}=-{\bm\gamma}_0 {\bf u}_{\lambda},
\label{pm}
\eea
where ${\bm\gamma}_0$ is calculated for $\epsilon=1$ and where we have taken the convention ${\bf w}_{-\lambda}={\bf w}_{\lambda}$ and $\hat{\bf w}_{-\lambda}=\hat{\bf w}_{\lambda}$.
Note that for $\epsilon=-1$,  $\tilde{\bf u}_\lambda$ and $\tilde{\bf v}_\lambda$ are the eigenstates corresponding to positive and negative eigenstates respectively.
The transformation defined  by Eq.($\ref{pm}$) can be interpreted as a transformation of particle in antiparticle of a network with the  opposite  orientation of the links  of the original network.}
Note however that this transformation does not hold for the harmonic eigenstates that for $\epsilon=-1$ read
\bea
\tilde{\bf u}_{0}={\bf v}_0,\quad
\tilde{\bf v}_0={\bf u}_0,
\eea
where $\tilde{\bf u}_0$ are the $\beta_1$ eigenstates corresponding to the eigenvalue $E=m_0>0$ and $\tilde{\bf v}_0$ are the $\beta_0$ eigenstates corresponding to eigenvalue $E=-m_0<0$ with ${\bf u}_0\neq {\bf v}_0$ as long as the network is not a linear chain.

The choice of the sign of $\epsilon$ is at this stage arbitrary, although it can change the physics of the problem if the network is not a closed one dimensional chain  which is self-dual. In particular we will show in the following that the choice of $\epsilon$ determines the value of the mass of the network.
Leaving this discussion for later on, for the moment we proceed our derivation assuming $\epsilon=1$ and hence $m>0$.

We consider now the second quantization of this field theory and hence we write $\bm\psi$ and $\bar{\bm\psi}$ as 
\bea
\bm\psi=\sum_{\lambda}\frac{1}{2E_{\lambda}}\left[b_\lambda{\bf u}_{\lambda}e^{-{\rm{i}}E_{\lambda} t}+c_{\lambda}^{\dag}{\bf v}_{\lambda}e^{{\rm{i}}E_{\lambda}t}\right],\nonumber \\
\bar{\bm\psi}=\sum_{\lambda}\frac{1}{2E_{\lambda}}\left[b_\lambda^{\dag}\bar{{\bf u}}_{\lambda}e^{{\rm{i}}E_{\lambda}t}+c_{\lambda}\bar{{\bf v}}_{\lambda}e^{-{\rm{i}}E_{\lambda}t}\right],
\label{psi_simple}
\eea
where  $\bar{\bf u}_\lambda={\bf u}_{\lambda}^{\dag}\bm\gamma_0$ and $\bar{\bf v}_{\lambda}={\bf v}_{\lambda}^{\dag}\bm\gamma_0$. Here $b^{\dag}_{\lambda}$ and $b_{\lambda}$ are the creation and annihilation operators of particles and $c^{\dag}_{\lambda}$ and $c_{\lambda}$ are creation and annihilation operator of the anti-particles obeying the anticommutation relations
\bea
\{b_{\lambda},b^{\dag}_{\lambda'}\}=2E_{\lambda}\delta_{\lambda,\lambda'},\nonumber \\
\{c_{\lambda},c^{\dag}_{\lambda'}\}=2E_{\lambda}\delta_{\lambda,\lambda'}.
\eea
Let us now consider the  Hamiltonian, given by
\bea
H=\bar{\bm\psi}{\rm i}\bm\gamma_0\partial_t\bm\psi.
\eea
Using the Eqs.(\ref{uno})  and (\ref{due}), it is easy to show that $H$ reads  
\bea
H=\sum_{\lambda>0}\frac{1}{2E_{\lambda}}E_{\lambda}[b^{\dag}_{\lambda}b_{\lambda}+c^{\dag}_{\lambda}c_{\lambda}],
\eea
upon redefining the zero of the energy.
The topological Dirac propagator is defined as
\bea
T(\bm\psi_{\alpha}(t),\bar{\bm\psi}_{\beta}(t'))=\theta(t-t')\bm\psi_{\alpha}(t)\bar{\bm\psi}_{\beta}(t')-\theta(t'-t)\bar{\bm\psi}_{\beta}(t'){\bm\psi}_{\alpha}(t).
\eea
Its expectation value on the ground state is given by 
\bea
\Avg{0|T(\bm\psi_{\alpha}(t),\bar{\bm\psi}_{\beta}(t'))|0}=\sum_{\lambda}[\theta(t-t'){\Theta}^+_{\alpha\beta}-\theta(t'-t){\Theta}^-_{\alpha\beta}]
\eea
where $\Theta^{\pm}_{\alpha\beta}$ read
\bea
\Theta^{+}_{\alpha\beta}=\sum_{\lambda}\frac{1}{2E_\lambda}[{\bf u}_{\lambda}\bar{\bf u}_{\lambda}]_{\alpha\beta}e^{-{\rm{i}}E_{\lambda}(t-t')},\nonumber \\
\Theta^{-}_{\alpha\beta}=\sum_{\lambda}\frac{1}{2E_\lambda}[{\bf v}_{\lambda}\bar{\bf v}_{\lambda}]_{\alpha\beta}e^{-{\rm{i}}E_{\lambda}(t'-t)}. 
\eea
Note that a  direct calculation of the exterior product ${\bf u}_{\lambda}\bar{\bf u}_{\lambda}$ and ${\bf v}_{\lambda}\bar{\bf v}_{\lambda}$ gives
\bea
{\bf u}_{\lambda}\bar{\bf u}_{\lambda}=\left(\begin{array}{cc}(E_{\lambda}+m){\bf w}_{\lambda}{\bf w}_{\lambda}^{\dag} &-\lambda {\bf w}_{\lambda}\hat{\bf w}_{\lambda}^{\dag}\\
\lambda \hat{\bf w}_{\lambda}{\bf w}_{\lambda}^{\dag}&(-E_\lambda+m) \hat{\bf w}_{\lambda}\hat{\bf w}_{\lambda}^{\dag}
\end{array}\right),\nonumber \\
{\bf v}_{\lambda}\bar{\bf v}_{\lambda}=\left(\begin{array}{cc}(E_{\lambda}-m){\bf w}_{\lambda}{\bf w}_{\lambda}^{\dag} &\lambda {\bf w}_{\lambda}\hat{\bf w}_{\lambda}^{\dag}\\
-\lambda \hat{\bf w}_{\lambda}{\bf w}_{\lambda}^{\dag}&(-E_\lambda-m) \hat{\bf w}_{\lambda}\hat{\bf w}_{\lambda}^{\dag}
\end{array}\right).
\eea

\subsection{The  mass of the network and the gap equation}
\label{sec:mass_net}
We consider now a massless topological Dirac field theory with a self-interaction term that is inspired by the Nambu-Jona-Lasinio model \cite{nambu1961dynamical}.
In particular we consider a discrete and topological  variation of the Nambu-Jona-Lasinio action in one dimension~\cite{thies2020phase,khunjua2017inhomogeneous,grabowska2013sign,khunjua2019charged}, hence involving only the gamma matrices $\bm\gamma_0$ and $\bm\gamma_5$.
Note however that the networks can have general topology, as the proposed theoretical framework does not constraint their topology in any way.

The action $\mathcal{S}$ of our discrete topological field theory can be expressed as 
\bea
\mathcal{S}=\mathcal{S}_0+\mathcal{S}_{int},
\eea
 with the unperturbed massless Dirac action $\mathcal{S}_0$  given by 
\bea 
\mathcal{S}_0=\bar{\bm\psi}(\rm{i}{\bm\gamma}_0\partial_t-{\bm\gamma}_0{\bf D})\bm\psi.
\eea
This corresponds to the massless ($m_0=0$)  action given by Eq.(\ref{S_simple}) with eigenstates corresponding to non-zero energy states related by chirality ${\bf v}_{\lambda}={\bm\gamma_0}{\bf u}_{\lambda}$ and ${\bf u}_{\lambda}={\bm\gamma_0}{\bf v}_{\lambda}$ given by 
\bea
{\bf u}_{\lambda}=\frac{1}{\sqrt{2}}\left(\begin{array}{c} {\bf w}_{\lambda} \\ \hat{\bf w}_{\lambda}\end{array}\right),\quad
{\bf v}_{\lambda}=\frac{1}{\sqrt{2}}\left(\begin{array}{c} {\bf w}_{\lambda}\\ -\hat{\bf w}_{\lambda}\end{array}\right).
\eea
The harmonic eigenstates $\lambda=E=0$ instead, are not related by chirality and are  given by 
\bea
{\bf u}_{0}=\left(\begin{array}{c} {\bf w}_{0} \\ 0\end{array}\right),\quad
{\bf v}_{0}=\left(\begin{array}{c} 0\\ \hat{\bf w}_{0}\end{array}\right).
\eea
In order to distinguish between the chiral component of $\bm\psi$ and the harmonic one, we adopt the notation 
\bea
\bm\psi=\bm\psi_0+\bm\psi_c,
\eea
with $\bm\psi$ given by Eq. (\ref{psi_simple}) and $\bm\psi_c$ given by 
\bea
\bm\psi_c=\sum_{\lambda>0}\frac{1}{2E_{\lambda}}\left[b_\lambda{\bf u}_{\lambda}e^{-{\rm{i}}E_{\lambda} t}+c_{\lambda}^{\dag}{\bf v}_{\lambda}e^{{\rm{i}}E_{\lambda}t}\right].
\eea
Let us now introduce the matrix $\bm\gamma_5$ given by 
\bea
{\bm\gamma}_5=\sum_{\lambda>0}[{\bf u}_\lambda{\bf u^{\dag}}_{\lambda}-{\bf v}_{\lambda}{\bf v}^{\dag}_{\lambda}],
\eea
which in matrix form reads
\bea
{\bm\gamma}_5=\sum_{\lambda>0}\left(\begin{array}{cc}0&{\bf w}_{\lambda}\hat{\bf w}^{\dag}_{\lambda}\\
\hat{\bf w}_{\lambda}{\bf w}^{\dag}_{\lambda}&0\end{array}\right).
\eea
The matrix $\gamma_5$ can be used to define the projected states
\bea
\bm\psi_R=\frac{1}{2}(1+\bm\gamma_5)\bm\psi_c,
\nonumber \\
\bm\psi_L=\frac{1}{2}(1-\bm\gamma_5)\bm\psi_c,
\eea
which are eigenstates of the chiral operator.
As in the continuous case the unperturbed action $\mathcal{S}_0$
 can be expressed in terms of $\bm\psi_R$ and $\bm\psi_L$ as 
 \bea
\mathcal{S}_0=\bar{\bm\psi}_R\bm\gamma_0(\rm{i}\partial_t-{\bf D})\bm\psi_R+\bar{\bm\psi}_L\bm\gamma_0(\rm{i}\partial_t-{\bf D})\bm\psi_L,
\eea
which can be shown to be invariant for  $U(1)$ and for chiral transformations as well, i.e. for transformations of the type 
\bea
\bm\psi\to e^{\rm{i}\alpha}\bm\psi,\label{U1}\\
\bm\psi\to e^{\rm{i}\beta \bm\gamma_5}\bm\psi. \label{chiral_trans}
\eea
Note that under the chiral transformation Eq.(\ref{chiral_trans}) we have
\bea
\left(\begin{array}{c}\bar{\bm\psi}_c\bm\psi_c\\
{\rm{i}}\bar{\bm\psi}_c{\bm\gamma}_5\bm\psi_c\end{array}\right)\to\left(\begin{array}{cc}\cos(\beta)&\sin(\beta)\\
-\sin(\beta)&\cos(\beta)\end{array}\right) \left(\begin{array}{c}\bar{\bm\psi}_c\bm\psi_c\\
{\rm{i}}\bar{\bm\psi}_c{\bm\gamma}_5\bm\psi_c\end{array}\right).
\label{chiral_trans2}
\eea
We now consider the self-interaction by introducing the interacting action $\mathcal{S}_{int}$ inspired by the Nambu-Jona-Lasinio model\cite{nambu1961dynamical}:
\bea 
\mathcal{S}_{int}=\frac{ {\tilde{g}}}{\bar{N}}[(\bar{\bm\psi}\bm\psi)^2-(\bar{\bm\psi}{\bm\gamma}_5\bm\psi)^2],
\eea
where $\bar{N}=N_0+N_1$ and where $ {\tilde{g}}>0$ is the coupling constant.
Let us observe that  $\bar{\bm\psi}\bm\psi$ and $\bar{\bm\psi}{\bm\gamma}_5\bm\psi$ can be written as 
\bea
\bar{\bm\psi}\bm\psi=\bar{\bm\psi}_0\bm\psi_0+\bar{\bm\psi}_c{\bm\psi}_c, \nonumber \\
\bar{\bm\psi}{\bm\gamma}_5\bm\psi=\bar{\bm\psi}_c{\bm\gamma}_5\bm\psi_c,
\eea
where
\bea
\bar{\bm\psi}_c\bm\psi_c=\bar{\bm\psi}_L{\bm\psi}_R+\bar{\bm\psi}_R{\bm\psi}_L,\nonumber \\
\bar{\bm\psi}_c{\bm\gamma}_5\bm\psi_c=
\bar{\bm\psi}_L{\bm\psi}_R-\bar{\bm\psi}_R{\bm\psi}_L.
\eea
We have therefore that the self-interaction term of the action, $\mathcal{S}_{int}$ can be written as 
\bea
\mathcal{S}_{int}=\mathcal{S}_{int,0}+\mathcal{S}_{int,c},
\eea
where 
\bea
\mathcal{S}_{int,0}=\frac{ {\tilde{g}}}{\bar{N}}(\bar{\bm\psi}_0\bm\psi_0)(\bar{\bm\psi}\bm\psi+\bar{\bm\psi}_c\bm\psi_c),\nonumber \\
\mathcal{S}_{int,c}=\frac{ {\tilde{g}}}{\bar{N}}[(\bar{\bm\psi}_c\bm\psi_c)^2-(\bar{\bm\psi}_c{\bm\gamma}_5\bm\psi_c)^2]=
4\frac{ {\tilde{g}}}{\bar{N}}(\bar{\bm\psi}_L{\bm\psi}_R)(\bar{\bm\psi}_R{\bm\psi}_L).
\eea
Note that while $\mathcal{S}_{int,c}$ is invariant under the chiral transformations given by  Eqs.(\ref{chiral_trans}), (\ref{chiral_trans2}), $\mathcal{S}_{int,0}$ is not chiral invariant, however for large $\bar{N}$ and small Betti numbers one can assume that this is a infinitesimal perturbation of the chiral symmetry.

By considering the mean-field, Hartree approximation, using similar steps used for the solution of the Nambu-Jona-Lasinio model \cite{nambu1961dynamical,hatsuda1994qcd}, the action reduces to the mean-field action $\mathcal{S}_{MF}$ given by 
\bea
\mathcal{S}_{MF}=\bar{\bm\psi}(\rm{i}{\bm\gamma}_0\partial_t-{\bm\gamma}_0{\bf D})\bm\psi-M(\bar{\bm\psi}\bm\psi)-\pi (\bar{\bm\psi}{\bm\gamma}_5\bm\psi),
\eea
where $M$ and $\pi$ can be put  self-consistently equal to
\bea
M=-2\frac{ {\tilde{g}}}{\bar{N}}\Avg{\bar{\bm\psi}\bm\psi},\quad \pi=\Avg{\bar{\bm\psi}{\bm\gamma}_5\bm\psi}=0.
\eea
Here $M$ describes the {\em mass of the network} that emerges from the self-interaction of the massless chiral topological Dirac field. The mass of the network that we have defined  breaks the chiral symmetry and corresponds to the  {\em Dirac mass}  of continuous field theories.
Specifically, we have that $M$ satisfies
\bea
M&=&-2\frac{ {\tilde{g}}}{\bar{N}}\Avg{\bar{\bm\psi}\bm\psi}
=2\frac{\tilde{g}}{\bar{N}} \lim_{t'\to t^{+}}\sum_{\alpha}\Avg{0|T(\bm\psi_{\alpha}(t),\bar{\bm\psi}_{\alpha}(t'))|0},
\eea
where the propagator is calculated for the topological Dirac field with mass $M$.
By using the explicit expression of the Dirac propagator for $\epsilon=1$ we then obtain the self-consistent {\em gap equation} for the mass $M=M_1$ given by
\bea
M_1= {\frac{g}{\bar{N}}}\beta_1 +\frac{g}{\bar{N}}\sum_{\lambda>0}\frac{M_1}{\sqrt{\lambda^2+M^2_1}},
\eea
where $ {g=2\tilde{g}}$.
This equation shows that the mass of the network is dependent on both the spectral properties of the network (through the spectrum $\lambda$) and the network topology through the dependence on the Betti number $\beta_1$.
The main difference with the self-consisted mass equation of the Nambu-Jona-Lasinio model is that the role of the bare mass  introduced by hand in the Nambu-Jona-Lasinio model has now a topological interpretation and is given by $g\beta_1/\bar{N}$.
Interestingly we observe that if $\epsilon=-1$ the equation for the mass of the network changes and we will obtain the mass $M=M_0$ given by 
\bea
M_0= {\frac{g}{\bar{N}}}\beta_0+\frac{g}{\bar{N}}\sum_{\lambda>0}\frac{M_0}{\sqrt{\lambda^2+M^2_0}},
\eea
where now the bare mass is  proportional to the Betti number $\beta_0$ instead of the Betti number $\beta_1$.
Hence the mass $M_1$ is the mass of the network corresponding to the choice $\epsilon=1$ while $M_0$ is the mass of the network corresponding to the choice $\epsilon=-1$.
As long as we consider a connected network, these two values of the mass are equal only for a closed $1$-dimensional chain for which we have $\beta_0=\beta_1$. Since for a tree we have $\beta_0=1$ and $\beta_1=0$, it follows that for a tree $M_1$ is given by 
\bea
M_1=\frac{g}{\bar{N}}\sum_{\lambda>0}\frac{M_1}{\sqrt{\lambda^2+M^2_1}},
\eea
and hence the non-zero solution for $M_1$ can be only observed for 
\bea
g>g_c=\left[\frac{1}{\bar{N}}\sum_{\lambda>0}\frac{1}{\lambda}\right]^{-1}.
\eea
Therefore  the mass $M_1$ displays a phase transition at $g=g_c$ with $M_1$ 
for $|g-g_c|\ll 1$  following the scaling behaviour
\bea
M_1=\left\{\begin{array}{cll}0&\mbox{if}& g<g_c\\
a(g_c-g)^{1/2}&\mbox{if}& g\geq g_c\end{array}\right.
\label{eq:scaling}
\eea
where $a>0$ is a constant.\\

 {We stress that  our framework applies to arbitrary networks. In particular the networks that we consider here  might strongly deviate not only from finite dimensional lattices but also from the wide class of discrete manifolds. Despite this, the  definition of the network mass that we have proposed can be related to the Dirac mass of the continuous $1+1$ Nambu-Jona-Lasinio models \cite{grabowska2013sign,kaplan1993chiral}.  
In the continuous field theories several additional theoretical aspects are also playing an important role.
On one side, for the continuous field theory in $1+1$ dimensions a chiral mass can be also defined in ways that does not break the chiral symmetry. Moreover for the continuous field theories in dimensions $2+1$ mass terms can be characterized by other discrete symmetries. These theoretical aspects of the continuous theory might possibly be relevant also in our discrete setting by restricting the focus to lattices instead of treating arbitrary networks. However these extensions are beyond the scope of this work and will possibly be the subject of future research. }
\begin{figure}[!htbp]
\begin{center}
\includegraphics[width=1.0\textwidth]{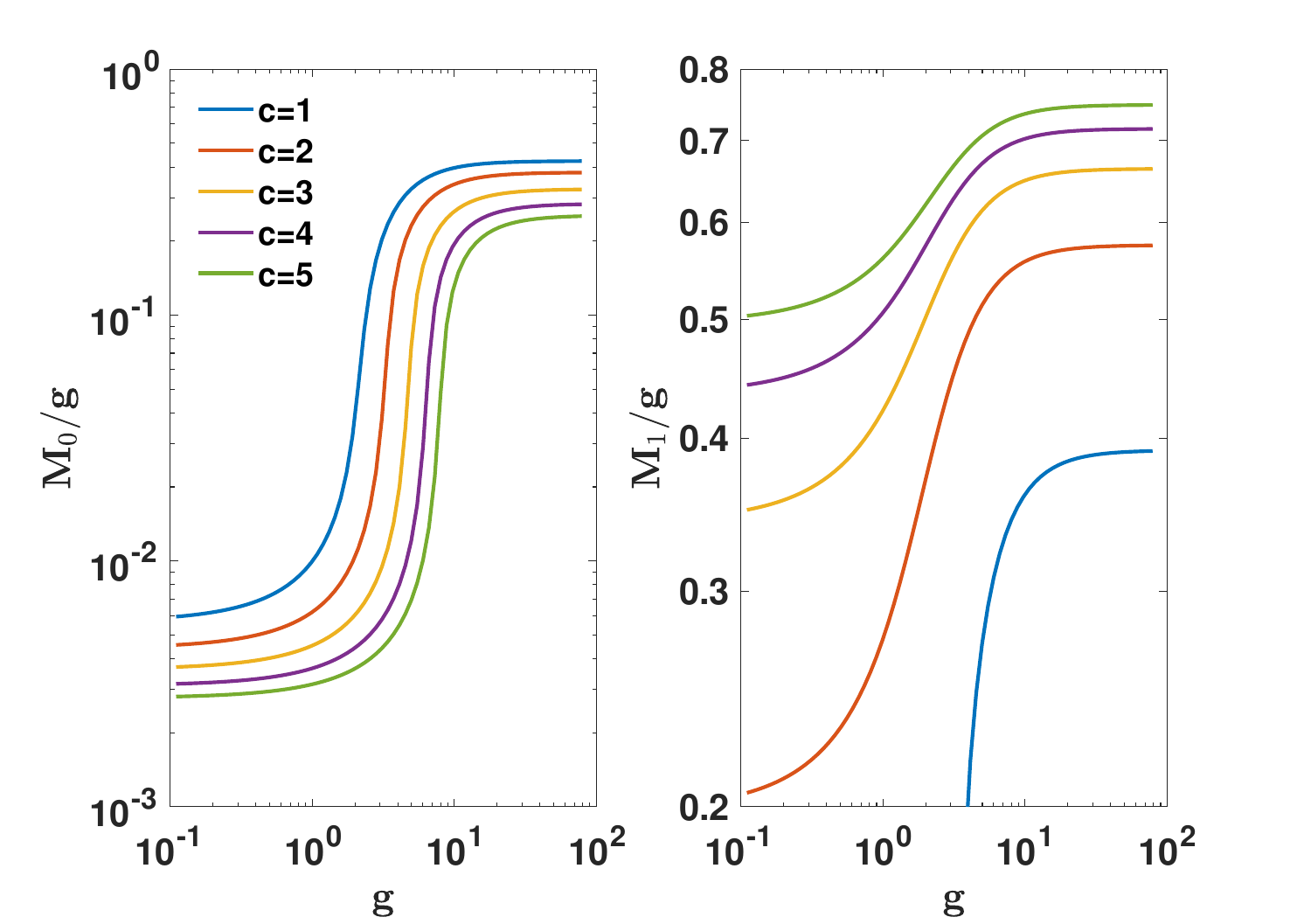}
\caption{{\bf The mass $M_0$ and $M_1$ of a Erd\"os-Renyi network.} The mass $M_0$ and $M_1$ of a Erd\"os-Renyi network of average degree $c$ and number of nodes $N_0=1000$ are plotted versus the coupling constant $g$. In this case both metric matrices ${\bf G}_0$ and ${\bf G}_1$ are taken to be identity matrices.}
\label{fig:ER}
\end{center}
\end{figure}
\begin{figure}[!htbp]
\begin{center}
\includegraphics[width=1.0\textwidth]{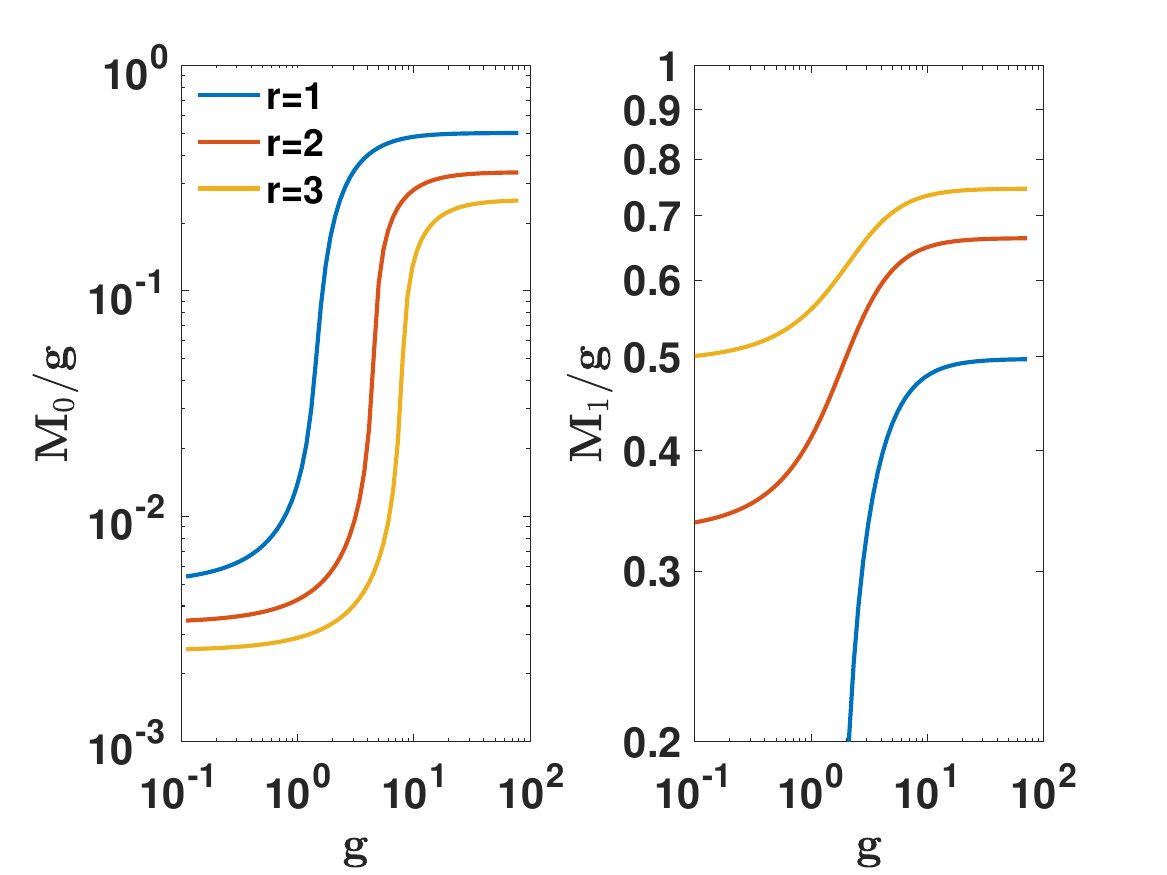}
\caption{{\bf The mass $M_0$ and $M_1$ of a Barabasi-Albert network.} The mass $M_0$ and $M_1$ of Barabasi-Albert network are plotted versus the coupling constant $g$.   The initial network is a clique of $r+1$ nodes, and each new node has  $r$ initial links. The total   number of nodes is $N_0=100$. In this case both metric matrices ${\bf G}_0$ and ${\bf G}_1$ are taken to be identity matrices.}
\label{fig:BA}
\end{center}
\end{figure}
\subsection{How the topology and geometry of networks affect the value of their mass}

In this section, we discuss numerical results evaluating the mass of different network models and real network data as well. Specifically, we consider the Erdős-Rényi (ER) networks \cite{ER}, the scale-free networks generated with the Barabási-Albert (BA) model \cite{BA}, the trees generated with the Bianconi-Barabási model \cite{bianconi2001bose} undergoing Bose-Einstein condensation on complex networks, and the real-weighted Pierre Auger scientific collaboration network~\cite{de2015identifying}.

\begin{figure}
\begin{center}
\includegraphics[width=1.0\textwidth]{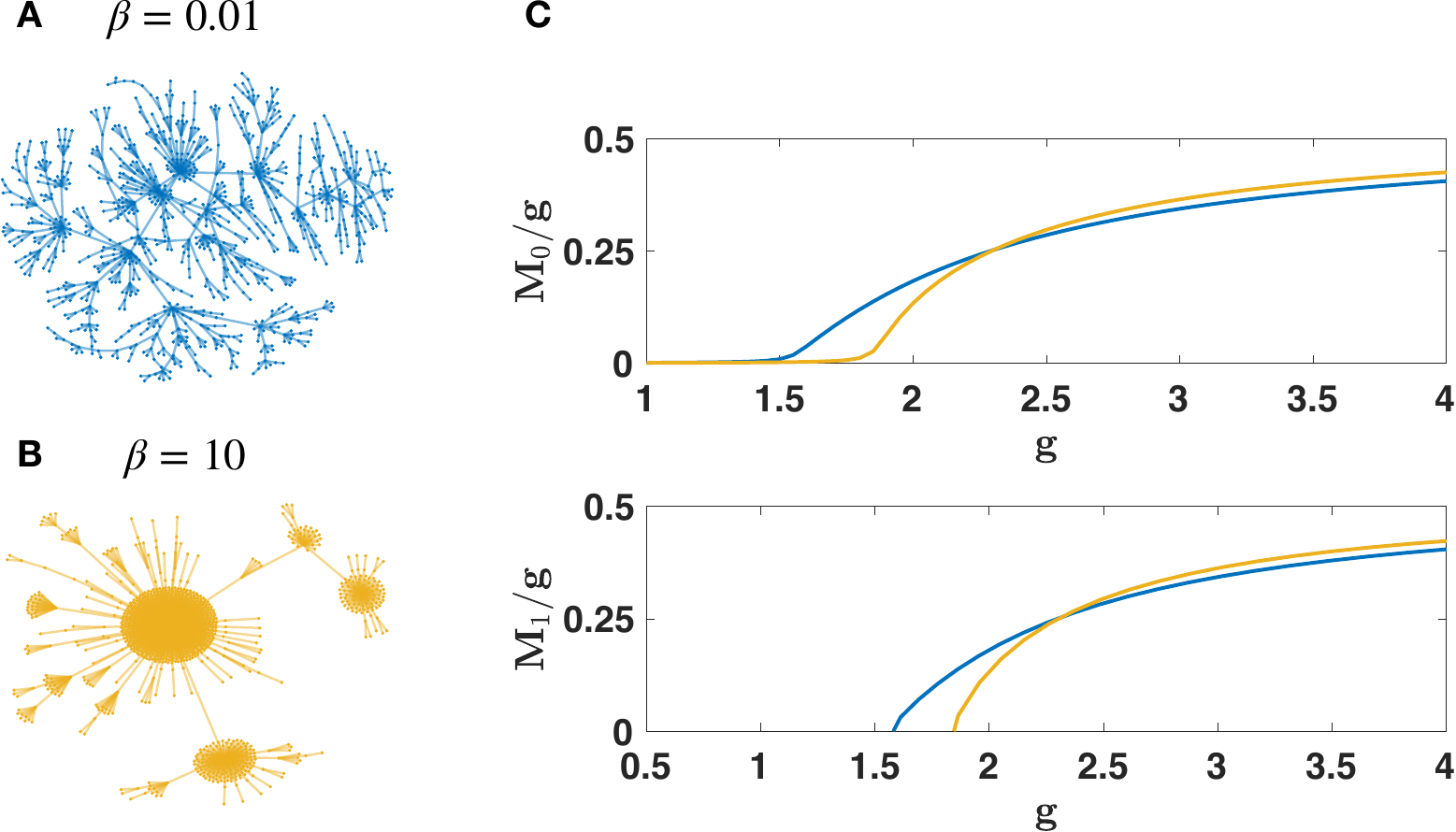}
\caption{{\bf The mass of the Bianconi-Barabasi model in the normal and in the condensed phase.} A realization of the Bianconi-Barabasi \cite{bianconi2001bose} tree is plotted for $\beta=0.01<\beta_c$ (normal phase, panel A) and for $\beta=10>\beta_c$ (condensed phase, panel B). The corresponding  mass $M_0$ and $M_1$ are plotted versus $g$ in panel C.  The number of nodes is $N_0=1000$. The parameter $\theta$ of the model is set to $\theta=2$.  In this case both metric matrices ${\bf G}_0$ and ${\bf G}_1$ are taken to be identity matrices.}
\label{fig:BB}
\end{center}
\end{figure}

The mass of (unweighted) ER networks (see Figure~\ref{fig:ER}) is shown to depend on the network average degree $c$ for both definitions of the mass $M_0$ and $M_1$. Let us observe that the average degree $c$ of ER networks modulates the expected number of connected components and the expected number of links, hence affecting the values of the Betti numbers $\beta_0$ and $\beta_1$. It follows that the dependence of $M_0$ and $M_1$ on the average degree $c$ implies their strong dependence on the topology of the random ER networks.
We also consider the mass of the unweighted BA model (see Figure~\ref{fig:BA}), which initially is formed by a clique of $r+1$ nodes, and in which each new node has degree $r$. These networks are always connected (i.e., $\beta_0=1$), and reduce to  trees for $r=1$. It follows that for $r=1$, the mass $M_1$ follows the scaling behaviour Eq. (\ref{eq:scaling}) for $g\simeq g_c$. Also in this case, we notice significant differences between the mass $M_0$ and the mass $M_1$ and strong dependence on $r$, which determines the Betti number $\beta_1$ of the network. Note that even if we consider exclusively trees, the network structure can vary widely, and these differences are reflected in the values of the network mass.
In order to show evidence for this effect, we focus on the Bianconi-Barabási model \cite{bianconi2001bose}, displaying the so-called Bose-Einstein condensation in complex networks as a function of the parameter $\beta$. Above this phase transition, in the network's normal phase, i.e., for $\beta<\beta_c$, the network is scale-free with a heterogeneous degree distribution, although all the nodes have an infinitesimal fraction of the links in the large network limit. Instead, in the condensed phase, i.e., for $\beta>\beta_c$, the network is dominated by hub nodes with a finite fraction of the links.
Here, we focus on trees generated with the Bianconi-Barabási (BB) model \cite{bianconi2001bose}, studied across the transition, and we show evidence that the mass $M_0$ and $M_1$ display clearly distinct functional behaviours as a function of $g$ above and below the phase transition (see Figure~\ref{fig:BB}). We note, in particular, that for both $n=0$ and $n=1$, the mass of the network $M_n$ is larger in the condensate phase than in the normal phase for high values of the coupling constant $g$, while it is smaller for low values of the coupling constant $g$.

The results obtained on random ER networks, on the BA networks, and on the BB networks show that the mass of the network strongly depends on the spectral and topological properties of the network.

\begin{figure}[!htbp]
\begin{center}
\includegraphics[width=1.0\textwidth]{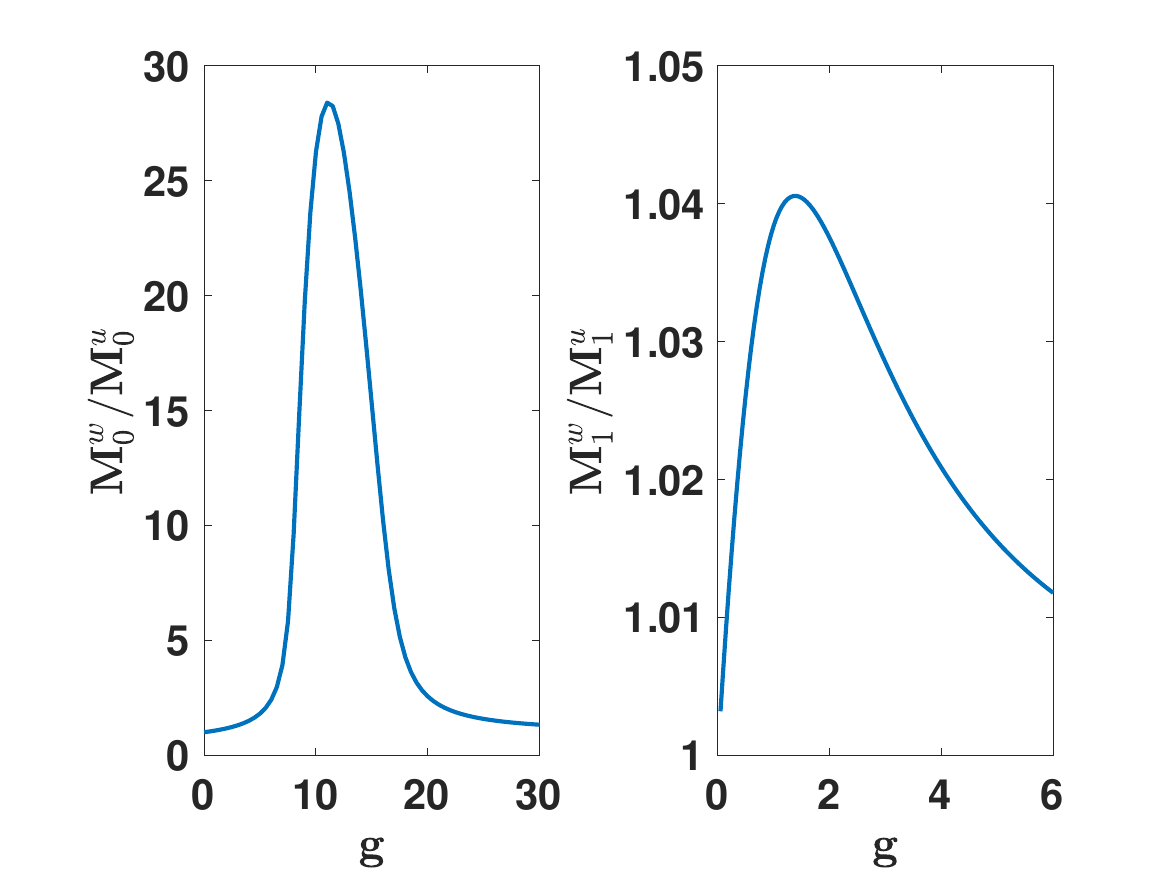}
\caption{{\bf The network mass depends on the weights of the links} We plot the ratio $M_0^{w}/M_0^{U}$ and $M_1^w/M_1^u$ between the weighted ($M_n^w$) and the unweighted ($M_n^u$) masses of the  Pierre Auger collaboration network~\cite{de2015identifying}
 (layer "Detector" with  $N_0=508$ nodes) versus the coupling constant $g$. The fact that these ratios are highly dependent on $g$ implies a non trivial effect of the metric degrees of freedom. Here we  use the metric matrix ${\bf G}_1^{-1}$ whose diagonal elements are the weights of the links, while  ${\bf G}_0$ is taken to be  the identity. Data available at the repository \cite{manlio_repository}.}
\label{fig:pierre_auger}
\end{center}
\end{figure}

By comparing the masses $M_n^{w}$ of the weighted Pierre Auger collaboration network (layer "Detector") and the masses $M_n^{u}$ of its unweighted version (see Figure~\ref{fig:pierre_auger}), we show that the weights, i.e., the metric degree of freedom, can also change the functional behaviour of the mass of the network versus the coupling constant $g$.
This implies that if the geometry and/or the topology of the network is allowed to change the mass of the topological Dirac field will also change. Therefore possibly one could envisage to formulate a coupled dynamics between the network structure and the topological Dirac field theory.

\section{The mass of higher-order networks}
\label{sec:higher}
In this Section  we move from networks to higher-order networks \cite{bianconi2021higher} providing a generalization of the theoretical framework that allows us to define  the {\em mass of a simplicial complex}. Note that also in this case, strictly speaking this term indicates the mass of a topological Dirac field defined on the simplicial complex.
\subsection{Higher dimensional topological Dirac operator}
For notational convenience  we will discuss the case of a $2$-dimensional simplicial complex. The generalization to higher-dimensional simplicial complexes is also briefly discussed at the end of this section. 

We consider a $2$ dimensional simplicial formed by  $N_0$ nodes, $N_1$ links and $N_2$ triangles. Every simplex (nodes, links, triangles) is oriented with an orientation induced by the nodes labels. 
On the $2$-dimensional simplicial complex it is possible to define $0,1$ and $2$ cochains, where the $0$, and $1$ cochains are  defined as in the case of networks. Similarly the  $2$-cochains are functions defined on each $2$-simplex (filled triangle) of the simplicial complex, where among $2$-cochains we define a scalar product given by the $L_2$ norm.
The nodes and links are weighted as for the case of simple networks. Additionally every triangle $[ijk]$ is associated a positive weight $w_{[ijk]}$.
The discrete curl is the function $\delta_2:C^1\to C^2$ such that 
if $g=\delta_2f$ then ${\bf g}={\bar{\bf B}}_2{\bf f}$  with the elements $g_{[ijk]}$ given by
\bea
g_{[ijk]}=\sqrt{{w}_{[ijk]}}\left(\frac{f_{[ij]}}{\sqrt{w_{[ij]}}}+\frac{f_{[jk]}}{\sqrt{{w}_{[jk]}}}-\frac{f_{[ik]}}{\sqrt{w_{[ik]}}}\right).
\eea 
On the considered $2$-dimensional simplicial complex we  define the topological spinor $\bm\psi\in C^0\oplus C^1\oplus C^2$ as   the direct sum of $0,1$ and $2$ cochains. 
  In matrix form the topological spinor reads
\bea
\bm\psi=\left(\begin{array}{c} \bm\phi\\ \bm\xi\\ \bm \eta \end{array}\right).
\eea
where $\bm\phi\in C^0, \bm \xi\in C^1$ and $\bm\eta\in C^2$ or equivalently
\bea
\bm \phi=\left(\begin{array}{c} \phi_1\\ \phi_2\\\vdots \\\phi_{N_0}\end{array}\right), & \bm \xi=\left(\begin{array}{c} \xi_{\ell_1}\\ \xi_{\ell_2}\\\vdots\\ \xi_{\ell_{N_1}}\end{array}\right),\quad \bm \eta=\left(\begin{array}{c} \eta_{t_1}\\ \eta_{t_2}\\\vdots\\ \eta_{t_{N_2}}\end{array}\right),
\eea

On top of the topological spinor we define the discrete topological Dirac operator ${\bf D}:C^0\oplus C^1 \oplus C^2\to C^0 \oplus C^1 \oplus C^2$ that maps topological spinors to topological spinors.
The Dirac operator is defined  in matrix form as
\bea
\mathbf{D}&=&\left(\begin{array}{ccc}0& {\bf \bar{B}}_{[1]}^{\top}&0\\{\bf \bar{B}}_{[1]}& 0&{\bf \bar{B}}_{[2]}^{\top}\\0&{\bf \bar{B}}_{[2]}&0\end{array}\right),
\eea
where ${\bf \bar{B}}_{[n]}$ are the $n$th-coboundary matrices of dimension $N_{n}\times N_{n-1}$.
The coboundary matrices satisfy 
\bea
{\bf \bar{B}}_{[n+1]}{\bf \bar{B}}_{[n]}={\bf 0},\quad {\bf \bar{B}}_{[n]}^{\top}{\bf \bar{B}}_{[n+1]}^{\top}={\bf 0},
\label{bb}
\eea 
for any $n\geq 1$. \\
The Dirac operator is the ``square root" of the  Laplacian as we have 
\bea
{\bf D}^2={\bf\mathcal{L}}=\left(\begin{array}{ccc} {\bf {L}}_{[0]}& 0&0\\0&{\bf {L}}_{[1]}&0\\0&0&{\bf {L}}_{[2]}\end{array}\right)
\eea
where ${\bf {L}}_{[n]}$ indicate the $n$th weighted Hodge Laplacian operators with \bea
{\bf {L}}_{[0]}&=&{\bf \bar{B}}_{[1]}^{\top}{\bf \bar{B}}_{[1]},\nonumber \\{\bf {L}}_{[1]}&=&{\bf \bar{B}}_{[1]}{\bf \bar{B}}_{[1]}^{\top}+{\bf \bar{B}}_{[2]}^{\top}{\bf \bar{B}}_{[2]},\nonumber \\
{\bf {L}}_{[2]}&=&{\bf \bar{B}}_{[2]}{\bf \bar{B}}_{[2]}^{\top}.
\eea
Therefore the eigenvalues $\lambda$ of the Dirac operator are given by 
\bea
\lambda=\pm \sqrt{\mu}
\eea
where $\mu$ indicates any eigenvalue in the spectrum $\mathcal{L}$.
Let us define the two  operators
${\bf D}_{[1]}$ and ${\bf D}_{[2]}$ acting only on nodes and links and only on links and triangles respectively as
\bea
\mathbf{D}_{[1]}=\left(\begin{array}{ccc}0& {\bf \bar{B}}_{[1]}^{\top}&0\\{\bf \bar{B}}_{[1]}& 0&0\\0&0&0\end{array}\right),\quad \mathbf{D}_{[2]}=\left(\begin{array}{ccc}0& 0&0\\0& 0&{\bf \bar{B}}_{[2]}^{\top}\\0&{\bf \bar{B}}_{[2]}&0\end{array}\right)
\eea
A direct consequence of Eq.(\ref{bb}) is that the  Dirac operator ${\bf D}$ obeys Dirac decomposition which means
\bea
{\bf D}={\bf D}_{[1]}+{\bf D}_{[2]}
\eea
with 
\bea
\mbox{im}({\bf D}_{[1]})\subseteq \mbox{ker}({\bf D}_{[2]}),\\
\mbox{im}({\bf D}_{[2]})\subseteq \mbox{ker}({\bf D}_{[1]}).
\eea
Indeed this latter equation can be shown to hold because ${\bf D}_{[1]}{\bf D}_{[2]}={\bf D}_{[2]}{\bf D}_{[1]}={\bf 0}$ as it can be easily shown given Eq.(\ref{bb}).
It follows that the space of topological spinors be decomposed in a unique way into 
\bea
{\mathbb{C}}^{\bar{N}+N_2}=\mbox{im}({\bf D}_{[1]})\oplus \mbox{im}({\bf D}_{[2]})\oplus \mbox{ker}({\bf D}),
\eea
therefore the eigenvectors of the Dirac operator are either  in the image of ${\bf D}_{[1]}$ on in the image of ${\bf D}_{[2]}$ or are in the kernel  of ${\bf D}$ with  $\mbox{ker}({\bf D})=\mbox{ker}({\bf D}_{[1]})\cap\mbox{ker}({\bf D}_{[2]})$.

\subsection{Higher-order topological Dirac field theory}
We consider the higher-order topological Dirac field theory with the action 
\bea
\mathcal{S}=\bar{\bm\psi}[\bm\gamma_0({\rm{i}}\partial_t-{\bf D})-m_0]\bm\psi,
\label{S_higher}
\eea
where the matrix $\bm\gamma_0$ is now defined as
\bea
\bm \gamma_0=\epsilon\left(\begin{array}{ccc}{\bf I}& 0&0\\0& -{\bf I}&0\\0&0&{\bf I}\end{array}\right).
\eea
with $\epsilon\in \{1,-1\}$.
Note  that  in this higher-order setting we have the anticommutation relations
\bea
\{\bm\gamma_0,{\bf D}\}=0,\quad
\{\bm\gamma_0,{\bf D}_{[1]}\}=0,\quad\{\bm\gamma_0,{\bf D}_{[2]}\}=0.
\eea
By observing that this higher-order topological field theory decomposes into the sum of two field theories defined respectively on nodes and links and links/triangles, we have that for $\epsilon=1$ and $\lambda>0$ the positive energy eigenstates ${\bf u}_{\lambda}^{[n]}$ of energy $E=E_\lambda=\sqrt{\lambda^2+m^2}$ and the negative energy eigenstates ${\bf v}_{\lambda}^{[n]}$ of energy $E=-E_\lambda$ read 
\bea
{\bf u}_{\lambda}^{[1]}=\frac{1}{\sqrt{E_\lambda+m}}\left(\begin{array}{c} (E_{\lambda}+m){\bf w}_{\lambda}^{[1]}\\ {\lambda}\hat{\bf w}_{\lambda}^{[1]}\\0\end{array}\right),\nonumber \\
{\bf v}_{\lambda}^{[1]}=\frac{1}{\sqrt{E_{\lambda}+m}}\left(\begin{array}{c} -\lambda{\bf w}_{\lambda}^{[1]}\\ (E_{\lambda}+m)\hat{\bf w}_{\lambda}^{[1]}\\0\end{array}\right),\nonumber \\
{\bf u}_{\lambda}^{[2]}=\frac{1}{\sqrt{E_\lambda+m}}\left(\begin{array}{c} 0\\-(E_{\lambda}+m){\bf w}_{\lambda}^{[2]}\\ {\lambda}\hat{\bf w}_{\lambda}^{[2]}\end{array}\right),\nonumber \\
{\bf v}_{\lambda}^{[2]}=\frac{1}{\sqrt{E_{\lambda}+m}}\left(\begin{array}{c} 0\\\lambda{\bf w}_{\lambda}^{[2]}\\ (E_{\lambda}+m)\hat{\bf w}_{\lambda}^{[2]}\end{array}\right),
\eea
with ${\bf w}_{\lambda}^{[n]},\hat{\bf w}_{\lambda}^{[n]}$ being the right and left singular vectors of $\bar{\bf B}_{[n]}$ for $n\in \{1,2\}$ and $\lambda$ being their corresponding singular values.
For $\lambda=0$ we have instead
\bea
\hspace{-15mm}{\bf u}_{0}^{[1]}=\sqrt{2m}\left(\begin{array}{c} {\bf w}_{0}^{[0]}\\ 0\\0\end{array}\right),\quad
{\bf v}_{0}^{[1]}=\sqrt{2m}\left(\begin{array}{c} 0\\ {\bf w}_{0}^{[1]}\\0\end{array}\right),\quad
{\bf u}_{0}^{[2]}=\sqrt{2m}\left(\begin{array}{c} 0\\0\\ {\bf w}_{0}^{[2]}\end{array}\right),
\eea
where ${\bf w}_0^{[n]}$ indicate the eigenvectors of the $n$-th Hodge Laplacian ${\bf L}_{[n]}$ and where we put by convention ${\bf v}_{0}^{[2]}={\bf 0}$.
The second quantization of this field theory proceed similarly to the simple network case. In particular we will have $\bm\psi$ given by \bea
\bm\psi=\sum_{\lambda}\frac{1}{2E_{\lambda}}\left[b_\lambda^{[n]}{\bf u}_{\lambda}^{[n]}e^{-{\rm{i}}E_{\lambda} t}+\Big(c_{\lambda}^{[n]}\Big)^{\dag}{\bf v}_{\lambda}^{[n]}e^{{\rm{i}}E_{\lambda}t}\right],\eea
where the creation and annihilation operators follow the commutation relations 
\bea
\left\{b_{\lambda}^{[n]},\Big(b^{[n]}_{\lambda'}\Big)^{\dag}\right\}=2E_{\lambda}\delta_{\lambda,\lambda'},\nonumber \\
\left\{c_{\lambda}^{[n]},\Big(c^{[n]}_{\lambda'}\Big)^{\dag}\right\}=2E_{\lambda}\delta_{\lambda,\lambda'}
\eea
for $n\in \{1,2\}.$
\subsection{The mass of simplicial complexes}
We are now in the position to define the mass of a simplicial complex by generalizing the theory provided in Sec. \ref{sec:mass_net}.
To this end we consider the action 
\bea
\mathcal{S}=\mathcal{S}_0+\mathcal{S}_{int},
\eea
with $\mathcal{S}_0$ and $\mathcal{S}_{int}$ given by 
\bea
\mathcal{S}_0=\bar{\bm\psi}(\rm{i}{\bm\gamma}_0\partial_t-{\bm\gamma}_0{\bf D})\bm\psi, \nonumber \\
\mathcal{S}_{int}=\frac{ {\tilde{g}}}{\bar{N}}[(\bar{\bm\psi}\bm\psi)^2-(\bar{\bm\psi}{\bm\gamma}_5\bm\psi)^2],
\eea
where 
the matrix $\bm\gamma_5$ is now given by 
\bea
{\bm\gamma}_5=\sum_{n=1}^2\sum_{\lambda>0}\left[{\bf u}_\lambda^{[n]}\Big({\bf u}^{[n]}_{\lambda}\Big)^{\dag}-{\bf v}_{\lambda}^{[n]}\Big({\bf v}^{[n]}_{\lambda}\Big)^{\dag}\right],
\eea
and
\bea
\bar{N}=N_0+N_1+N_2.
\eea
The eigenstates of this massless field theory are given by for $\lambda\neq 0$ by 
\bea
{\bf u}_{\lambda}^{[1]}=\frac{1}{\sqrt{2}}\left(\begin{array}{c} {\bf w}_{\lambda}^{[1]} \\ \hat{\bf w}_{\lambda}^{[1]}\\0\end{array}\right),\quad
{\bf v}_{\lambda}^{[1]}=\frac{1}{\sqrt{2}}\left(\begin{array}{c} {\bf w}_{\lambda}^{[1]}\\ -\hat{\bf w}_{\lambda}^{[1]}\\0\end{array}\right)\nonumber \\
{\bf u}_{\lambda}^{[2]}=\frac{1}{\sqrt{2}}\left(\begin{array}{c} 0\\{\bf w}_{\lambda}^{[2]} \\ -\hat{\bf w}_{\lambda}^{[2]}\end{array}\right),\quad
{\bf v}_{\lambda}^{[2]}=\frac{1}{\sqrt{2}}\left(\begin{array}{c} 0\\{\bf w}_{\lambda}^{[2]}\\ \hat{\bf w}_{\lambda}^{[2]}\end{array}\right)
\eea
The harmonic eigenstates $\lambda=E=0$ instead, are not related by chirality and are  given by 
\bea
{\bf u}_{0}^{[1]}=\left(\begin{array}{c} {\bf w}_{0}^{[0]} \\ 0\\0\end{array}\right),\quad
{\bf v}_{0}^{[1]}=\left(\begin{array}{c} 0\\ {\bf w}_{0}^{[1]}\\0\end{array}\right),\quad {\bf u}_{0}^{[2]}=\left(\begin{array}{c} 0\\0\\{\bf w}_{0}^{[2]} \end{array}\right),
\eea
where ${\bf w}^{[n]}_0$ are the harmonic eigenvectors of the $n$-th Hodge Laplacian ${\bf L}^{[n]}$ and where for convention we put ${\bf v}_{0}^{[2]}={\bf 0}$.
By proceeding as in the network case, we can show that in the mean-field Hartree approximation the action $\mathcal{S}$ becomes 
\bea
\mathcal{S}_{MF}=\bar{\bm\psi}(\rm{i}{\bm\gamma}_0\partial_t-{\bm\gamma}_0{\bf D}-M)\bm\psi,
\eea
where $M$ is the {\em mass of the simplicial complex}.
For $\epsilon=1$ the mass is given by $M=M_1$  while for $\epsilon=-1$ the mass is given by $M=M_0$ with $M_n$ given by the gap equations
\bea
M_1= {\frac{g}{\bar{N}}}\beta_1 +\frac{g}{\bar{N}}\sum_{\lambda>0}\frac{M_1}{\sqrt{\lambda^2+M^2_1}}\nonumber \\
M_0= {\frac{g}{\bar{N}}}(\beta_0+\beta_2) +\frac{g}{\bar{N}}\sum_{\lambda>0}\frac{M_0}{\sqrt{\lambda^2+M^2_0}},
\eea
with  {$g=2\tilde{g}.$}
Through the Betti numbers both definitions of   mass $M_0$ and $M_1$ of simplicial complexes are strongly dependent on the simplicial complex topology. Moreover through the eigenvalues $\lambda$ both mass $M_0$ and $M_1$ are highly dependent on  simplicial complex topology and geometry.

A straightforward generalization of this definition to higher-dimensional simplicial complexes shows that in general we have two possible definitions of the mass of a simplicial complex with 
\bea
M_0= {\frac{g}{\bar{N}}}\sum_{k\geq 0}\beta_{2k}+\frac{g}{\bar{N}}\sum_{\lambda>0}\frac{M_0}{\sqrt{\lambda^2+M^2_0}}\nonumber \\
M_1= {\frac{g}{\bar{N}}}\sum_{k\geq 0}\beta_{2k+1} +\frac{g}{\bar{N}}\sum_{\lambda>0}\frac{M_1}{\sqrt{\lambda^2+M^2_1}}.
\eea
where $\lambda$ indicates the positive  eigenvalues of the higher-dimensional Dirac operator and $\bar{N}=\sum_{n=0}^d N_n$ where $N_n$ indicates the number of $n$-dimensional simplices.
We note that we have $M_0=M_1$ if and only if the Euler characteristic of the simplicial complex is zero.

\section{Conclusions}

Networks and simplicial complexes capture discrete topology and geometry. In the hypothesis that space-time is intrinsically discrete, a key question is to define a field theory on such structures. Here, we have used the discrete topological Dirac operator to define a self-interacting massless chiral action inspired by the Nambu-Jona-Lasinio model. We have shown that the chiral symmetry is broken, giving rise to a non-zero mass of the topological Dirac field that we call the network mass. This mass depends on the network structure, the network spectral properties, its metric degrees of freedom (weights of the links), and its topology (the Betti numbers).

Due to the fact that in this discrete setting, the matter-antimatter symmetry does not hold for harmonic eigenstates in general, two different definitions of the mass of a network can be given. The two definitions of the network mass differ from the value of their corresponding bare mass. This bare mass is here automatically determined by the topology of the network, i.e., by the Betti numbers of the network.

The mass of a simplicial complex can be defined in an analogous way and depends on the Betti numbers of the simplicial complex and the spectrum of the $d$-dimensional Dirac operator.

This work can be extended in several directions. First, although in the discrete setting, there is no need to introduce a regularizing cutoff, there might always be a dependence on the network size that could lead to interesting effects. For example, one could study how the mass of growing networks evolves over time. Secondly, it would be interesting to investigate the relation between the mass of networks derived from the unnormalized and the normalized Dirac operator (which has a bounded spectrum even in the large network limit). Secondly, the mass of a network could be defined using other algebras for the gamma matrices, for instance, investigating the corresponding definition of the mass on square $d$-dimensional lattices. Thirdly, one could mimic quantum chromodynamics by considering discrete topological Dirac fields of different flavor.
Finally one could launch a new research direction aimed at investigating  the coupled dynamics of the topological Dirac field,  and the evolution of the network structure (including its geometry and/or its topology). This coupled dynamics  might lead to interesting phenomena possibly including topological phase transitions of the network.

In conclusion, we hope that this work could provide a fertile ground for research at the interface between network theory and quantum gravity~\cite{bianconi2017emergent,bianconi2016network,wu2015emergent,chen2013statistical,he2020graph,akara2021birth,kleftogiannis2022emergent,Astridnet,reitz2020higher,Severini,Trugenberger1,Trugenberger2,kleftogiannis2022physics} and, more generally, quantum mechanics~\cite{bianconi2001bose,anand2011shannon,biamonte2019complex,de2016spectral,bottcher2022complex,tian2023structural,nokkala2020probing,nokkala2016complex,nokkala2018reconfigurable}, which has received increasing attention recently. Moreover, the hereby defined mass of simple and higher-order networks could prove itself as a relevant metric to characterize simple and higher-order networks coming from interdisciplinary applications.

\section*{Data Availability}
The Pierre Augier collaboration network data is published in ~\cite{de2015identifying} and freely available in the repository \cite{manlio_repository}.

\section*{Code Availability}
All the codes used in this work are available upon request.

\end{document}